\documentclass[a4paper,12pt]{article}
\usepackage{mathrsfs}
\usepackage{graphicx} 
\usepackage{epstopdf}

\begin{document}

\hoffset = -1truecm \voffset = -2truecm \baselineskip = 10 mm

\title{The gluon condensation effect in the cosmic hadron spectra}

\author{Wei Zhu$^1$, Pen Liu$^1$, Jianhong Ruan$^1$, Ruiqin Wang$^2$ and Fan Wang$^3$
\\
\normalsize $^1$Department of Physics, East China Normal University,
Shanghai 200241, China \\
\normalsize $^2$ School of Physics and Physical Engineering, Qufu
Normal University,
Shandong 273165, China \\
\normalsize $^3$Department of Physics, Nanjing University,
Nanjing,210093, China\\
}

\date{}

\newpage

\maketitle
\begin{abstract}

       Hardening of cosmic proton- and nuclei-spectra is explained by using
the gluon condensation (GC) model, which states that a large amount
of gluons in proton may condense near the high energy threshold. The
results present the GC-effect as
common origin of a series of anomalous astrophysical phenomena
including the broken power-law in gamma-ray spectra, the excess in
positron- and electron-spectra and hardening of proton- and
nuclei-spectra.

\end{abstract}

{\bf keywords}:  cosmic ray theory-gamma ray theory-ultra high energy cosmic rays

\vskip 1truecm

\newpage
\begin{center}
\section{\bf Introduction}
\end{center}

    The spectra of cosmic hadrons (pion, nucleon and nuclei) above
energy $E\sim 10GeV$ per nucleon are thought to be the result of
acceleration mechanism in supernova remnants (SNRs) and they present
a simple power-law except at "knee" ($E\sim 10^3~ TeV$) and "ankle"
($E\sim 10^6~ TeV$). The power-law of energy spectrum is a general
rule of the cosmic-ray (CR) spectra at high energy. It is described
by a straight line of the energy spectrum with a fixed index in the
double-logarithmic coordinator. However, the precise measurements
show that these CR spectra at rigidity $\tilde{R}>300~GV$ exhibit a
remarkable hardening with increasing energies [1]. In particular,
proton and helium spectra present an extra peak at $E\sim 11~TeV$.

The origin of the spectral hardening has several explanations: it
could be due to the energy dependence of the diffusion coefficient
[2,3]; the possible nonlinear acceleration in the source [4,5] and
the existence of extra local sources inside SNRs [6,7]. The origin
of the CR spectral hardening is still an open question in CR physics
[8-10].

The gluons inside proton dominate the proton collisions at high
energy and their distributions obey the evolution equations based on
Quantum Chromodynamics (QCD). A QCD analysis shows that the
evolution equations will become nonlinear due to the initial gluons
correlations at high energy and it results in a chaotic solution
beginning at a threshold critical energy [11-13]. Most surprisingly,
the dramatic chaotic oscillations produce strong shadowing and
antishadowing effects, they may converge gluons to a state at a
critical momentum. This is the gluon condensation (GC) in proton. In
this work we propose that the GC-effect of hadronic collisions in
SNR breaks the power-law of the hadron flux and leads to the
observed hardening in cosmic hadronic spectra.

The GC means a lot of gluons accumulate at a critical momentum, and
the secondary particles are significantly increased near the
GC-threshold. Proton-proton (or nucleus-nucleus) collisions are
general events inside SNR, which produce lots of secondary particles
(electron-positron, pion, proton-antiproton, nuclei...). Therefore,
the GC-effect should induce the characteristic excess in CR spectra
originated from $p-p$ collisions, provided the GC-threshold
$E_{p-p}^{GC}$ enters the observable high energy region. In our
previous work [14] we successfully used the GC-model to explain two
seemingly completely different events of the CR spectra: the excess
in positron flux and the broken power-law in gamma-ray spectrum of a
SNR. We have shown that the excess in the CR positron spectrum
observed by AMS originates mainly from the GC-effect in Tycho's
supernova remnant. A following naive idea is that the similar excess
should also occur in cosmic hadron spectra, since the secondary
hadrons are the main products in high energy proton-proton
collisions.

We find that the GC-effect may cause a large amount of herium-4 to
build up in the formation processes of proton-proton collisions. It
directly leads to the excess nucleons and nuclei peaked at $E\sim
m_{He}/m_{\pi}\times 400~GeV\simeq 11~TeV$, $400~GeV$ is a observed
broken point of gamma ray from Tycho's SNR. The GC-effect also
predicts that the strength of nuclei fluxes sensitively relates to
the average binding energy of the nucleus in the SNR environment.
Thus, a series of interesting anomalous power-law show their
intrinsic connection through the GC-effect.

We will discuss the properties of the GC effect at Sec.2. Then in
Sec. 3 we predict the proton spectrum. Based on the above results,
hardening of the cosmic nuclei spectra are studied in Sec. 4. The
discussions and a summary are given in Sec. 5.

\newpage
\begin{center}
\section{\bf The GC Model}
\end{center}

The secondary particles in CRs may origin from the hadronic
processes $p(A)+p(A)\rightarrow \pi^{\pm}+\pi^0+p+\bar {p}+others$.
Let us first consider a normal high-energy heavy-ions collisions
without the GC-effect. It is generally accepted that the partons
(quarks and gluons) in the central region of the collision have gone
through the following thermalization process: (I) all partons
convert to constituent quarks (CQs); (II) all constituent quarks
combine to pion, nucleon (proton and neutron), anti-nucleon..., they
form a dense and hot hadronic cluster: the fireball; (III) the
fireball expands rapidly and the hadron density decreases rapidly.
In this cooling process part of nucleons and anti-nucleons may form
nuclei and anti-nuclei. That is

$$I(G, q \overline{q}\rightarrow CQs)$$
$$\rightarrow II (CQs\rightarrow\pi,p,n...)$$
$$\rightarrow III(p,n,\overline{p},\overline{n}
\rightarrow A,\overline{A}).\eqno(2.1)$$

     Now we consider a high energy
collision of heavy ions with the GC-effect inside SNR. One can
expect that the CQ cluster will be produced in the central region of
the collision as similar to the above normal relativistic heavy ion
collision. However, a large number of gluons will be piled in the
central region with the collision energy larger than the
GC-threshold $E_{A-A}>E_{A-A}^{GC}$. A new research [15,16] found
that CQs are formed after bare quark absorbs gluons. Therefore,
there are enough CQs to form maximum number of hadrons due to the
GC-effect. All available relative kinetic energy of the constituents
is almost used to construct the hadrons in the central region. We
call it as the cold ball. This assumption described successfully the
broken power-law in CR gamma fluxes, if these hadrons are pions
[17]. In this work we further consider the contributions of nucleons
and anti-nucleons, which are accompanied by pions in a narrow phase
space. In order to get the maximum nucleon occupancy in a small
phase space, these nucleons and anti-nucleons should be in
the Boson-configuration in the cold ball due to Pauli principle.
Helium-4 ($^4He$) and anti-Heilum-4 ($^4\overline{He}$) are the
condensable and more stable Bosons comparing with deuteron.
Therefore, the cold ball is quite different from the fireball. We
have pointed out in our previous paper [17] that
$\pi^++\pi^-\rightarrow 2\pi^0$ since
$m_{\pi^+}+m_{\pi^-}>2m_{\pi^0}$ and then $\pi^0\rightarrow
2\gamma$. Thus, there is a large number of photons surrounding $^4He$, then
$^4He$ is quite possible to be dissolved by photons, i.e.,
$^4He(\gamma,D)D, ^4He(\gamma, p)T$ and $^4He(\gamma,n)^3He$. We
only consider the first reaction since its product is the
Boson-configuration, which can accommodate more hadrons in a small
phase space. Thus, we have the following sub-processes
$^4He\rightarrow D+D;~~D\rightarrow p+n.$

Then part of these dense secondary nucleons and antinucleons
recombine light nuclei. Note that most of antinucleons are
annihilated by nucleons within SNR, therefore, we don't consider the
antiproton spectrum in this report. Besides, the light nuclei with
small average binding energy are easily decomposed during
collisions. The above descriptions about the cosmic hadron spectra
in SNR with the GC-effect can be summarized as the following nuclear
reaction chain

$$ I'(condesated~G, q \overline{q}\rightarrow maximum~CQs)$$
$$\rightarrow II' (maximun~CQs\rightarrow maximum~ \pi^0,^4He)$$
$$\rightarrow
II''(decomposed~^4He~by~\gamma\rightarrow dense ~p,n,\overline{p},
\overline{n})$$
$$\rightarrow III'(dense~ p,n,\overline{p},
\overline{n} \rightarrow A,\overline{A}).\eqno(2.2)$$

Now we can easily generalize the GC-model for the lepton flux, which
has been derived in [14] to include the proton and nuclei fluxes. We
denote that $N_{\pi}(E_{p-p},E_{\pi})$ and $N_{He}(E_{p-p},E_{He})$
as $\pi^0$- and $^4He$-numbers with energies $E_{\pi}$ and $E_{He}$;
$E_{p-p}$ is the energy of incident proton in the rest frame of
target. According to the above discussions, we can directly write
the relativistic invariant and energy conservation as
$$(2m_p^2+2E_{p-p}m_p)^{1/2}\simeq E_{p1}^*+E_{p2}^*+N_{\pi}m_{\pi}+N_{He}m_{He}, \eqno(2.3)$$
$$E_{p-p}+m_p\simeq m_{p1}\gamma_1+m_{p2}\gamma_2+[N_{\pi}m_{\pi}+N_{He}m_{He}]\gamma,\eqno(2.4)$$
here $E_{p1}^*$ and $E_{p2}^*$ are the energy of the two secondary
leading protons in the CM system and we have considered $m_p\ll
E_{p-p}$.

 Denote

$$E_{He}/E_{\pi}=E_{He}^{GC}/E_{\pi}^{GC}=m_{He}/m_{\pi}\equiv\zeta,\eqno(2.5)$$ and

$$N_{He}/N_{\pi}\equiv\eta\ll 1,\eqno(2.6)$$ where we use $\eta=0.01$
to estimate the ratio of baryon and meson numbers. Note that the
value of $\eta$ is not well defined here. However, its correction to
the following parameters in Equations (2.13)-(2.16) can be neglected
if $\eta$ varies over a wide range. We rewrite Equations (2.3) and
(2.4) as
$$(2m_p^2+2E_{p-p}m_p)^{1/2}=E_{p1}^*+E_{p2}^*+N_{\pi}m^* \eqno(2.7)$$
$$E_{p-p}+m_p=m_{p1}\gamma_1+m_{p2}\gamma_2+N_{\pi}m^*\gamma, \eqno(2.8)$$where
$m^*=(1+\eta\zeta)m_{\pi}$. We take the inelasticity $K\sim 0.5$
[18] and set
$$E_{p1}^*+E_{p2}^*=(\frac{1}{K}-1)N_{\pi}m^*,\eqno(2.9)$$
and
$$m_{p1}\gamma_1+m_{p2}\gamma_2=(\frac{1}{K}-1)N_{\pi}m^*\gamma,\eqno(2.10)$$
One can  get the solutions $N_{\pi}(E_{p-p},E_{\pi})$ (or
$N_{He}(E_{p-p},E_{He})$) for $p-p$ collisions in the $GeV$-unit
$$\ln N_{\pi}=0.5\ln E_{p-p(A)}+a_{\pi}, ~~\ln N_{\pi}=\ln E_{\pi}+b_{\pi}, \eqno(2.11)$$and
$$\ln N_{He}=0.5\ln E_{p-p(A)}+a_{He}, ~~\ln N_{He}=\ln E_{He}+b_{He}. \eqno(2.12)$$
The parameters $a_{\pi}$, $b_{\pi}$, $a_{He}$ and $b_{He}$ are
$$a_{\pi} \equiv 0.5\ln (2m_p)-\ln m^*+\ln K,\eqno(2.13)$$
$$b_{\pi} \equiv \ln (2m_p)-\ln m^*m_{\pi}+\ln K,\eqno(2.14)$$
$$a_{He} \equiv 0.5\ln (2m_p)-\ln m^*+\ln K+\ln{\eta},\eqno(2.15)$$
$$b_{He} \equiv \ln (2m_p)-\ln m^*m_{He}+\ln K+\ln{\eta}. \eqno(2.16)$$
Equations (2.11) and (2.12) give the one-to-one relations among
$N_{\pi}$, $N_{He}$ and $E_{p-p}$, which lead to the
GC-characteristic spectra.

\newpage
\begin{center}
\section{\bf Cosmic proton spectrum}
\end{center}

    According to the GC-model in Sec.2, cosmic proton flux reads
$$\Phi_p(E)=\Phi^0_p(E)+\Phi^{GC}_p(E),\eqno(3.1)$$where $\Phi^0_p(E)$ is the background spectrum
and $\Phi^{GC}_p(E)$ is the GC-contributions through the hadronic
chain $p+p\rightarrow^4He$, $^4He\rightarrow 2D$ and $D\rightarrow
p+n$,

$$\Phi_p^{GC}(E)=C_p\left(\frac{E}{E_{He}^{GC}}\right)^{-\beta_p}
\int_{E}dE_D\left(\frac{E_D}{E_{He}^{GC}}\right)^{-\beta_D}
\int_{E_{He}^{min}}^{E_{He}^{max}}
dE_{He}\left(\frac{E_{p-p}}{E_{p-p}^{GC}}\right)^{-\beta_{p'}}$$
$$\times N_{He}(E_{p-p},E_{He})\frac{d\omega_{He-D}(E_{He},E_D)}{dE_D}
\frac{d\omega_{D-p}(E_D, E)}{dE}$$
$$=C_p\left(\frac{E}{E_{He}^{GC}}\right)^{-\beta_p}\int_{E}\frac{dE_D}{E_D}
\left(\frac{E_D}{E_{He}^{GC}}\right)^{-\beta_D}
\int_{E_{He}^{GC}~or~E_D}^{E_{He}^{max}}
dE_{He}\left(\frac{E_{p-p}}{E_{p-p}^{GC}}\right)^{-\beta_{p'}}N_{He}(E_{p-p},E_{He})
\frac{2}{\beta_{He}E_{He}}$$
$$=\left\{
\begin{array}{ll}
\frac{2C_pe^{b_{He}}}{2\beta_{p'}-1}E_{He}^{GC}\left(\frac{E}{E_{He}^{GC}}\right)^{-\beta_p}
\left[\frac{1}{\beta_D}
\left(\frac{E}{E_{He}^{GC}}\right)^{-\beta_D}+
(\frac{1}{\beta_D+2\beta_{p'}-1}-\frac{1}{\beta_D})\right]
 {~~~~\rm if~} E\leq E_{He}^{GC}\\\\
\frac{2C_pe^{b_{He}}}{(2\beta_{p'}-1)(\beta_D+2\beta_{p'}-1)}
(E_{He}^{GC})\left(\frac{
    E}{E_{He}^{GC}}\right)^{-\beta_p-\beta_D-2\beta_{p'}+1}
{~~~~~~~~~~~~~~~~~~~~~~\rm if~} E>E_{He}^{GC}
\end{array} \right., \eqno(3.2)$$

where the integral lower-limit  takes $E_{He}^{GC}$ (or $E_D$) if
$E_{He}\leq E_{He}^{GC}$ (or if $E_{He}> E_{He}^{GC}$). $-\beta_p$
is the suppression index of secondary particle p during propagation,
and $-\beta_{p'}$ is the index of incident proton which may carry
information about their origin. A factor
$(E_{He}/E_{He}^{GC})^{-\beta_{He}}$ has been incorporated into
$(E_{p-p}/E_{p-p}^{GC})^{-\beta_{p'}}$.

Note that comparing with the high energy $E_{p-p}$, we neglect the
mass of deuterium. After taking average over possible directions of
deuterium, the energy distribution of deuterium is equal
probability, i.e., the normalized spectrum is
$$\frac{d\omega_{He-D}(E_{He},E_D)}{dE_D}=\frac{1}{E_{He}}.\eqno(3.3)$$
Thus, the processes $p+p\rightarrow He\rightarrow D\rightarrow p+n$
are similar to $p+p\rightarrow \pi^0\rightarrow \gamma\rightarrow
e^++e^-$ The later has been discussed as a hadronic mechanism of the
positron spectrum in our previous work [17,19], it is written as

$$\Phi_e^{GC}(E)=C_e\left(\frac{E}{E_{\pi}^{GC}}\right)^{-\beta_e}
\int_{E}dE_{\gamma}\left(\frac{E_{\gamma}}{E_{\pi}^{GC}}\right)^{-\beta_{\gamma}}
\int_{E_{\pi}^{min}}^{E_{\pi}^{max}}
dE_{\pi}\left(\frac{E_{p-p}}{E_{p-p}^{GC}}\right)^{-\beta_p}$$
$$\times N_{\pi}(E_{p-p},E_{\pi})\frac{d\omega_{\pi-\gamma}(E_{\pi},E_{\gamma})}{dE_{\gamma}}
\frac{d\omega_{\gamma-e}(E_{\gamma}, E)}{dE}$$
$$=C_e\left(\frac{E}{E_{\pi}^{GC}}\right)^{-\beta_e}\int_{E}\frac{dE_{\gamma}}{E_{\gamma}}
\left(\frac{E_{\gamma}}{E_{\pi}^{GC}}\right)^{-\beta_{\gamma}}
 \int_{E_{\pi}^{GC}~or~E_{\gamma}}^{E_{\pi}^{max}}
dE_{\pi}\left(\frac{E_{p-p}}{E_{p-p}^{GC}}\right)^{-\beta_p}N_{\pi}(E_{p-p},E_{\pi})
\frac{2}{\beta_{\pi}E_{\pi}}$$
$$=\left\{
\begin{array}{ll}
\frac{2C_ee^{b_{\pi}}}{2\beta_p-1}E_{\pi}^{GC}\left(\frac{E}{E_{\pi}^{GC}}\right)^{-\beta_e}
\left[\frac{1}{\beta_{\gamma}}
\left(\frac{E}{E_{\pi}^{GC}}\right)^{-\beta_{\gamma}}+
(\frac{1}{\beta_{\gamma}+2\beta_p-1}-\frac{1}{\beta_{\gamma}})\right]
 {~~~~~~~~~~~\rm if~} E\leq E_{\pi}^{GC}\\\\
\frac{2C_ee^{b_{\pi}}}{(2\beta_p-1)(\beta_{\gamma}+2\beta_p-1)}
(E_{\pi}^{GC})\left(\frac{
    E}{E_{\pi}^{GC}}\right)^{-\beta_e-\beta_{\gamma}-2\beta_p+1}
{~~~~~~~~~~~~~~~~~~~~~~~~~~~\rm if~} E>E_{\pi}^{GC}
\end{array} \right. \eqno(3.4)$$

A strong prediction of the GC-model is that the secondary proton and
positron spectra are closely related. Equations (3.2) and (3.4) show
that the peak positions $E_{He}^{GC}$ and $E_{\pi}^{GC}$ of two
kinds of spectra are related by equation (2.5), i.e.,

$$E_{He}^{GC}=\frac{m_{^4He}}{m_{\pi}}E_{\pi}^{GC}. \eqno(3.5)$$
Besides, $\beta_{p'}$ in Equation (3.2) is equal to $\beta_p$ in
Equation (3.4). The $E_{He}^{GC}$ is fixed to be 10.8~TeV using
Equation (3.5) and $\beta_{p'}=1.4$ in Equation (3.2). Note that we
do not investigate the origin of the background, which is taken an
extension line in Figure 1 (dashed line). The GC-effect superimposed
on the background results a broken power-law in the proton spectrum
as shown by solid curve in Fig. 1. The predicted position of an
extra peak at $E\sim 10.8 TeV$ is consistent with the data. One can
find that hardening beginning from kinetic energy $\sim 300~GeV/n$
is the slope part of the excess peak due to the GC-effect.

\newpage
\begin{center}
\section{\bf Cosmic nuclei spectra}
\end{center}

    The accelerator experiments show that the production yield of
light-nuclei decreases on the order of magnitude as the nuclear mass
increment one by one. The production mechanism of light nuclei in
the field of heavy-ion physics is one open questions [20]. Normally,
the nucleus is randomly combined by nucleons with different
distributions in an inflated fireball. Unfortunately, astronomical
measurements cannot get these distributions in SRN. We can only use
an ideal model to build the formation process of light nuclei at the
GC-environment. According to the GC model, all nucleons are closely
co-moving with a large common velocity. Although the temperature is
very low in the cold ball and the kinetic energy of the nucleon
seems not enough to overcome the Coulomb potential, however, the
nucleons are closely co-moving, their wave functions have a larger
probability of overlap during flight inside a GC-source.  We
introduce the following simple formula to describe the relation
between the detected nuclear flux $\Phi_A$ and nucleon flux $\Phi_p$

$$\frac{\Phi_{A}(E_A)}{\Phi_p(E_p)}= R_A^{A-1},\eqno(4.1)$$where
$\Phi_A$ has a same form as $\Phi_p$ but $E_A\simeq AE_p$ if
neglecting the binding corrections at the recombination process. The
parameter $R_A$ is defined as the recombination probability that two
nucleons combine to one nucleus. In this work, $R_A$ is a free
 parameter extracted from data.

   Similar to Equation (3.1) we write

$$\Phi_A(E)=\Phi^0_A(E)+\Phi^{GC}_A(E),\eqno(4.2)$$and

$$\Phi^{GC}_A(E)=R_A^{A-1}\Phi^{GC}_p(E).\eqno(4.3)$$That is

$$\Phi^{GC}_A(E)=R_A^{A-1}$$
$$\times\left\{
\begin{array}{ll}
3.7\times 10^{-7}\left(\frac{E}{10800GeV}\right)^{-1.23}\\
-2.9\times10^{-7}\left(\frac{E}{10800GeV}\right)^{-0.73}
,{~\rm if~} E\leq 10800GeV\\\\
8\times 10^{-8}\left(\frac{
    E}{10800GeV}\right)^{-3.03},{~\rm if~} E>10800GeV.
\end{array} \right. \eqno(4.4)$$Using the data [21,22]
of fluxes for helium and Equation (4.4) we get $R_A=0.46$ for $^4He$
in Figure 2, where the background is taken as an extension line of
$\Phi^0_A$. Note that the $He$-flux was then treated as containing
only $^4He$ [23].

In the following Figures 3-8, we present the fluxes for lithium,
beryllium, boron, carbon, oxygen, nitrogen and oxygen multiplied by
$\tilde{R}^{2.75}$ as functions of rigidity $\tilde{R}$. The
relation of kinetic energy per nucleon $E_k$ and rigidity is
$E_k=(\sqrt{Z^2\tilde{R}^2+m_A^2}-m_A)/A$, where $m_A$ is the
nuclear mass and $Z$ is the nuclear charge.

    The values of $R_A$ are 0.46, 0.14, 0.19, 0.34, 0.51, 0.51 and 0.60 for He, Li, Be, B, C, N and O.
Interestingly, the above $R_A$-order is similar to that of average
binding energy of these nuclei (see Figure 9). We suggest that
$\Phi_A$ will be obviously reduced if the nucleus $A$ has a smaller
binding energy, since it is easy to be decomposed during they pass
through the matter inside SNR. A smaller binding energy implies a
larger decomposition, which corresponding to a smaller value of
$R_A$. Therefore, it is different from Equation (4.1). We cannot
record deuterium-, tritium and helium3-fluxes in the CRs at the
energy band of the GC-effect since their average binding energies
are much smaller than other nuclei. On the other hand, as
mentioned in Sec. 2, $\Phi_A\gg \Phi_{\overline{A}}$ for cosmic nuclear fluxes due to strong absorption of the
SRN matter.
\newpage
\begin{center}
\section{Discussion and summary}
\end{center}

   Based on a QCD predicted GC-effect, we assume that $p-p$ and $A-A$ collisions
produce a lot of hadrons, they fill a limited phase space with the
largest number of particles, it forms a cold ball, which is consists
of $\pi^0$ and $^4He$. Then $^4He$ again decomposes to nucleons by
photons. Part of these nucleons recombine into different light-nuclei
according to Equation (4.1). Although there is no direct evidence
for the above series of assumptions, the predicted eight spectra in
Figures 1-8 can be tested by experiments in the near future.

We emphasize that these nuclei spectra present a series of
properties arising from the GC-effect. (i) the extra GC-source in a
neighboring SRN produces strong proton and nuclei fluxes, which
break the CR power-law; (ii) the peak positions of exceeded hadrons
are fixed by Equation (3.5) due to the GC-effect in Equations (2.7)
and (2.8); (iii) the GC-effect leads to co-moving nucleons, they are
recombined into nuclei according to Equation (4.1).

The GC opens a new window for us to understand a series of anomalous
excesses in the high energy CR spectra from the quark-gluon level.
Some of the topics is current interests, for example, the typical
broken power-law in gamma spectra [17,19], the excess in positron-,
electron-spectra [14], even in proton and helium, and the hardening
light nuclei-spectra in this work, they may originate from a same
basic process: a large amount of gluons may condense near the high
energy threshold. Combining Figures 1-8, we predict hardening in a
serious of light nuclei spectra. The above anomalous power-law
originates from the GC-effect in the nucleon collisions inside SRNs.

  The GC-effect forms a new physical state in the universe: high density but cold hadronic
cluster, which is quite different from the fireball in heavy ion
collision and even in big bang nucleosynthesis (BBN). We think that
this is a new field that has not been known to us so far.

    In summary, a QCD research predicts that gluons in protons may condensed at a
critical momentum in high energy collision of proton-proton. The
condensed gluons produce a special hadronic state, in which pion and
$^4He$ have high densities. The photons are radiated by neutral
pions and decompose helium into nucleons. Part of these nucleons
recombine to light nuclei. They are superimposed on the background
flux to form hardening in cosmic proton and nuclei spectra. We find
that hardening beginning from rigidity $\tilde{R}\sim 300 GV$ is the
slope part of the excess peak near $E\sim 11TeV$ due to the
GC-effect. The above explanation is also consistent with the
observed the broken power-law in gamma ray spectra of a neighboring
SNR and the anomalous excess in CR positron-electron spectra.
Because gamma, lepton, nucleon and nuclei are the products of high
energy proton-proton collisions and their strength is governed by
gluon distribution in the proton, the GC-effect should affect all
these spectra with a uniform law. Our results support this
prediction.

\noindent {\bf ACKNOWLEDGMENTS} We thank F.L. Shao and J. Song for
useful discussions about the recombination mechanism. This work is
supported by the National Natural Science of China (No.11851303).

\newpage

\begin{figure}
    \begin{center}

 \includegraphics[width=0.8\textwidth]{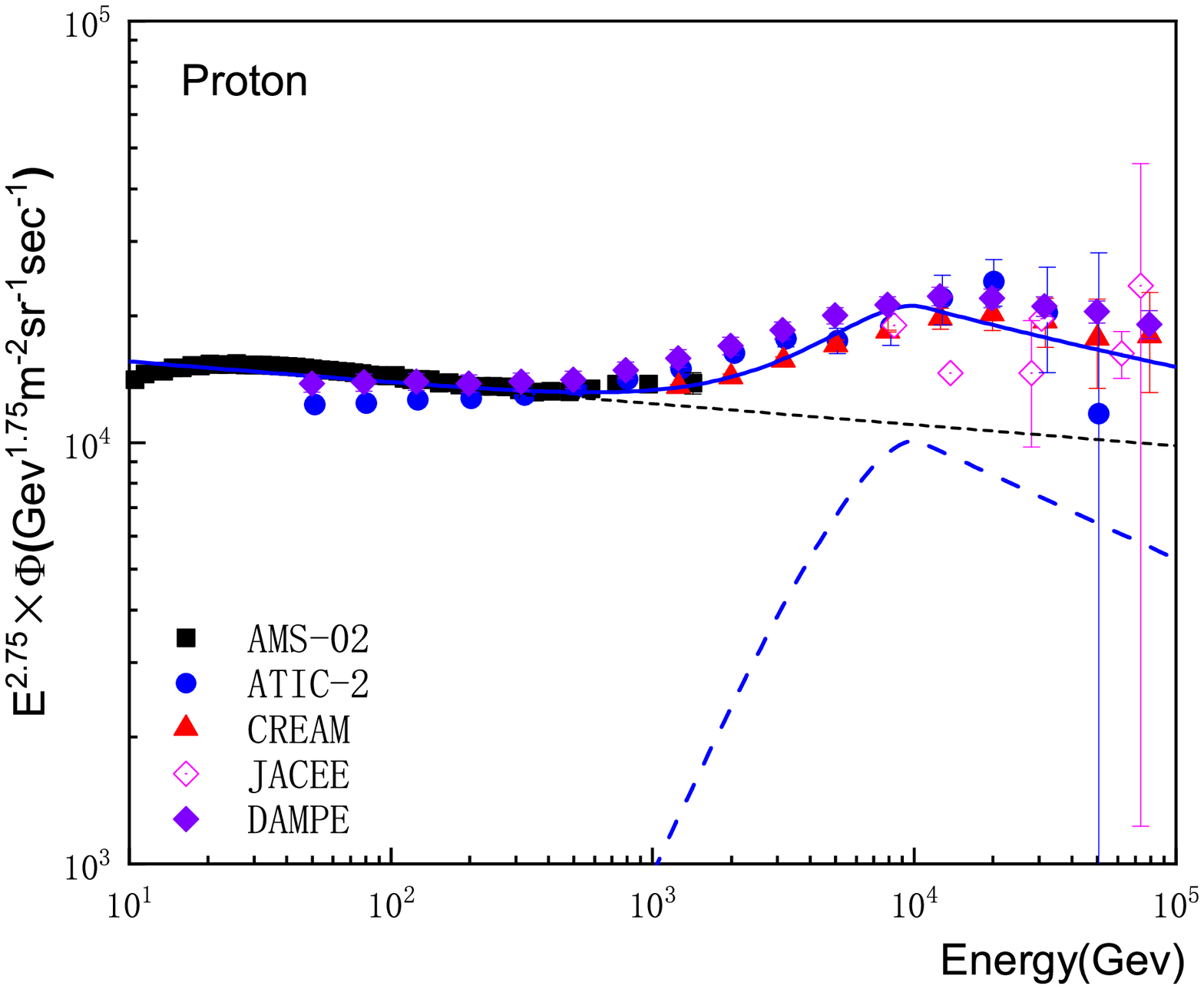}
        \caption{\label{fig:fig1} The proton energy spectrum multiplied by
$E^{2.75}$ as function of proton energy (solid curve). The dotted
curve is background. The GC-contributions (dashed curve) are
predicted by Equation (3.2). Note that the peak position of proton
flux is fixed by that of positron flux and Equation (3.5). The used
parameters are $\beta_p=0.73$, $\beta_D=0.5$ and $C_p=1.55\times
10^{-11} (GeV^{-3}\cdot m^{-2}\cdot sr^{-1}\cdot sec^{-1})$. The
position of the peak $E_{He}^{GC}$ in Equation (3.2) is fixed by the
positron spectrum. The data are taken from AMS [24,25], ATIC [26],
CREAM [21], JACEE [27], RUNJOB [22] and DAMPE [28].}\label{Fig.1}

    \end{center}
\end{figure}

\begin{figure}
    \begin{center}

 \includegraphics[width=0.8\textwidth]{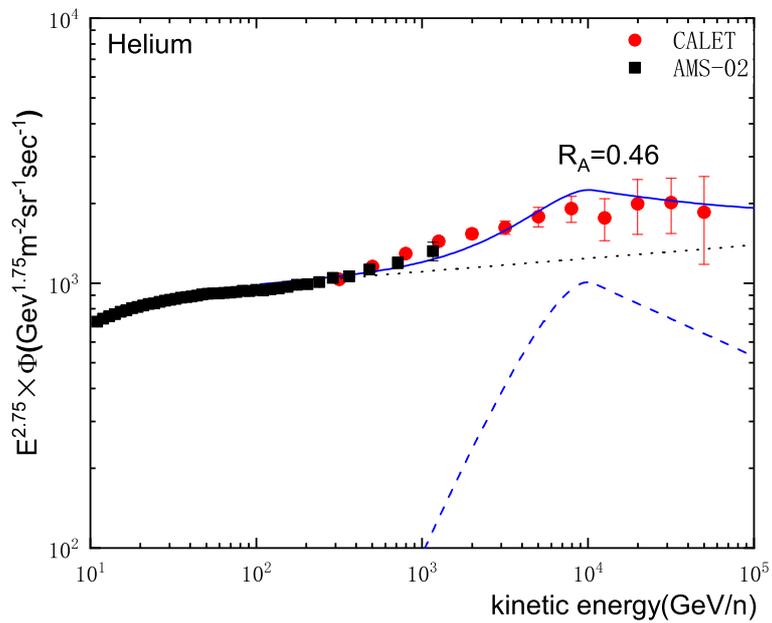}
        \caption{\label{fig:fig2} The helium energy spectrum multiplied by
$E^{2.75}$ as function of proton energy (solid curve). The dotted
curve is background. The GC-contributions (dashed curve) are
predicted by Equation (4.4). The $E_k$ position of the peak is fixed
by the positron spectrum. A free parameter $R_A=0.46$. The data are
taken from AMS-02 [25] and CREAM [21].}\label{Fig.2}

\end{center}
\end{figure}

\begin{figure}
\begin{center}

 \includegraphics[width=0.8\textwidth]{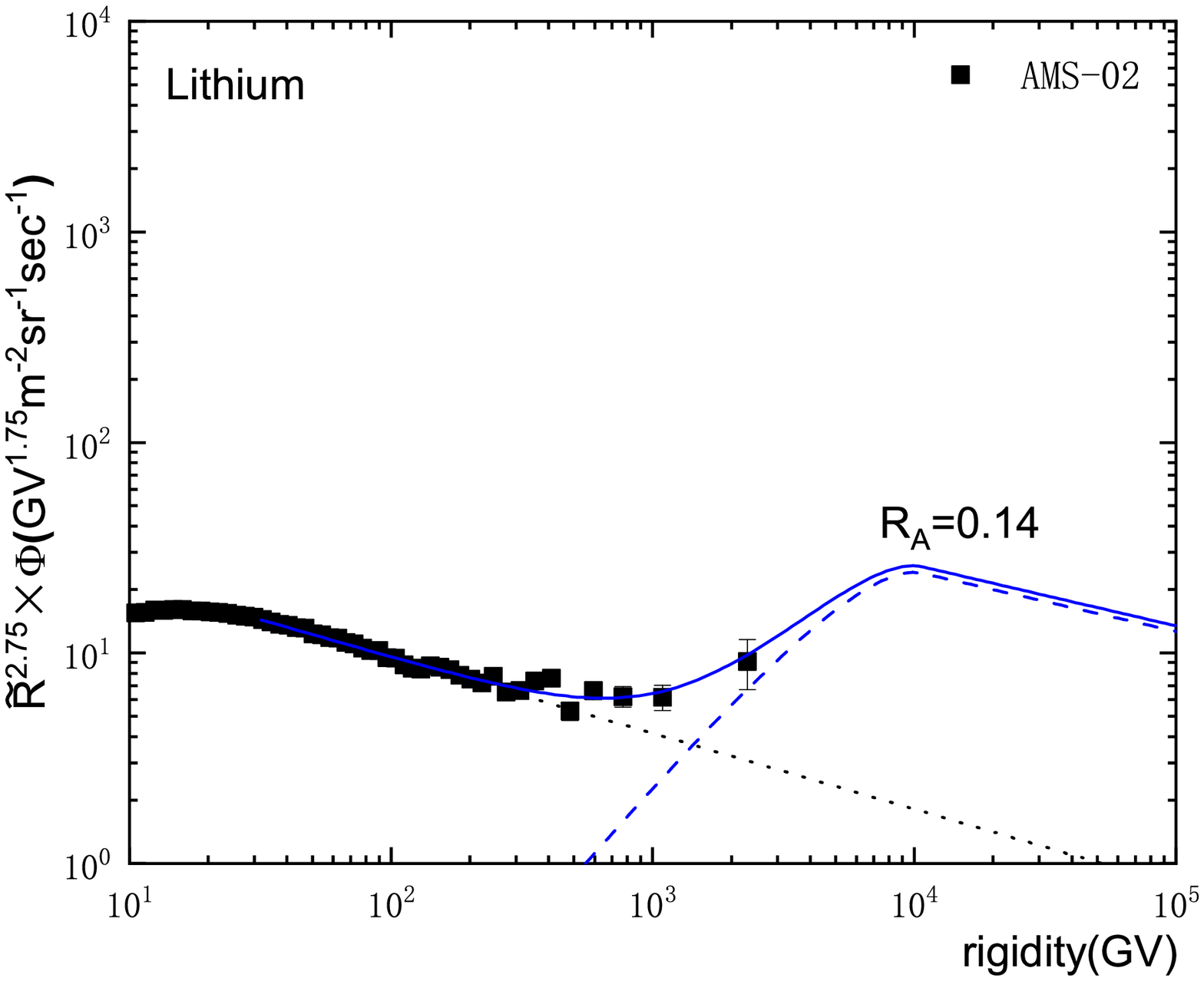}
        \caption{\label{fig:fig3} The lithium energy spectrum multiplied by
$\tilde{R}^{2.75}$ as function of rigidity (solid curve). The dotted
curve is background. The GC-contributions (dashed curve) are
predicted by Equation (4.4). The $\tilde{R}$ position of the peak is
fixed by the positron spectrum. A free parameter $R_A=0.14$. Data
are taken from the AMS-02 [29].}\label{Fig.3}

    \end{center}
\end{figure}

\begin{figure}
    \begin{center}

 \includegraphics[width=0.8\textwidth]{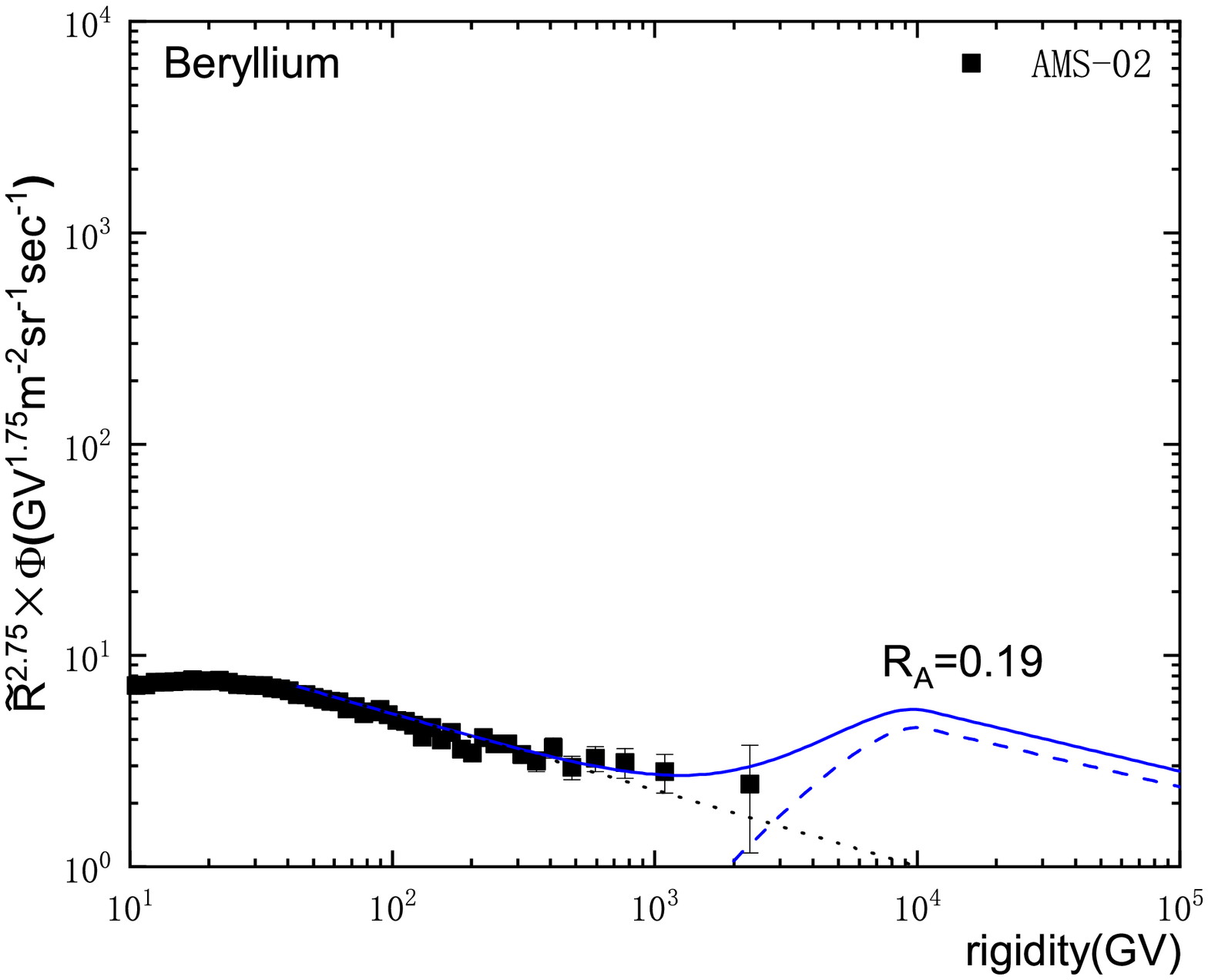}
        \caption{\label{fig:fig4} The beryllium energy spectrum multiplied by
$\tilde{R}^{2.75}$ as function of rigidity (solid curve). The dotted
curve is background. The GC-contributions (dashed curve) are
predicted by Equation (4.4). The $\tilde{R}$ position of the peak is
fixed by the positron spectrum. A free parameter $R_A=0.19$. Data
are taken from the AMS-02 [29].} \label{Fig.4}

\end{center}
\end{figure}

\begin{figure}
    \begin{center}

 \includegraphics[width=0.8\textwidth]{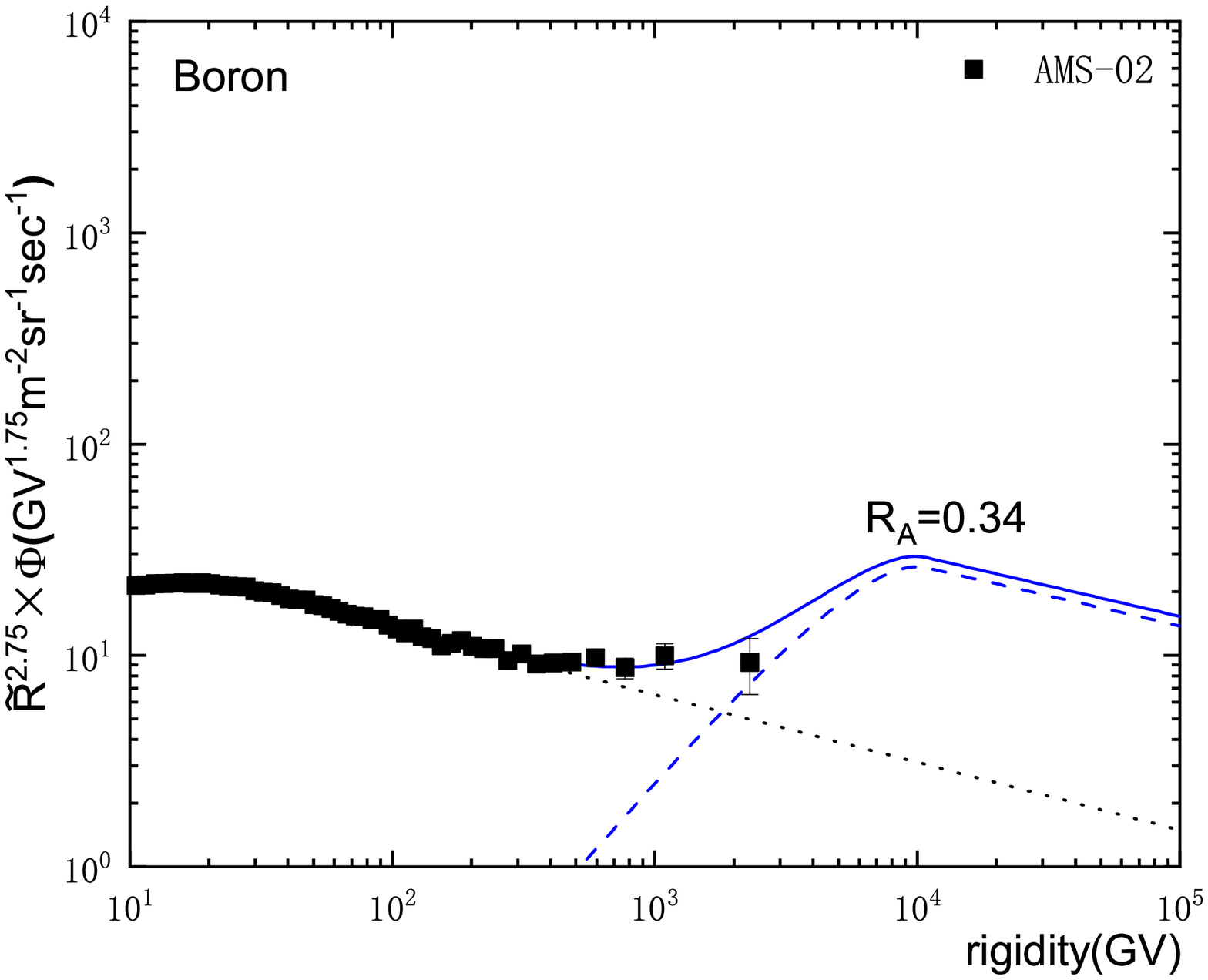}
        \caption{\label{fig:fig5} The boron energy spectrum multiplied by
$\tilde{R}^{2.75}$ as function of rigidity (solid curve). The dotted
curve is background. The GC-contributions (dashed curve) are
predicted by Equation (4.4). The $\tilde{R}$ position of the peak is
fixed by the positron spectrum. A free parameter $R_A=0.34$. Data
are taken from the AMS-02 [29].}\label{Fig.5}

\end{center}
\end{figure}

\begin{figure}
    \begin{center}

 \includegraphics[width=0.8\textwidth]{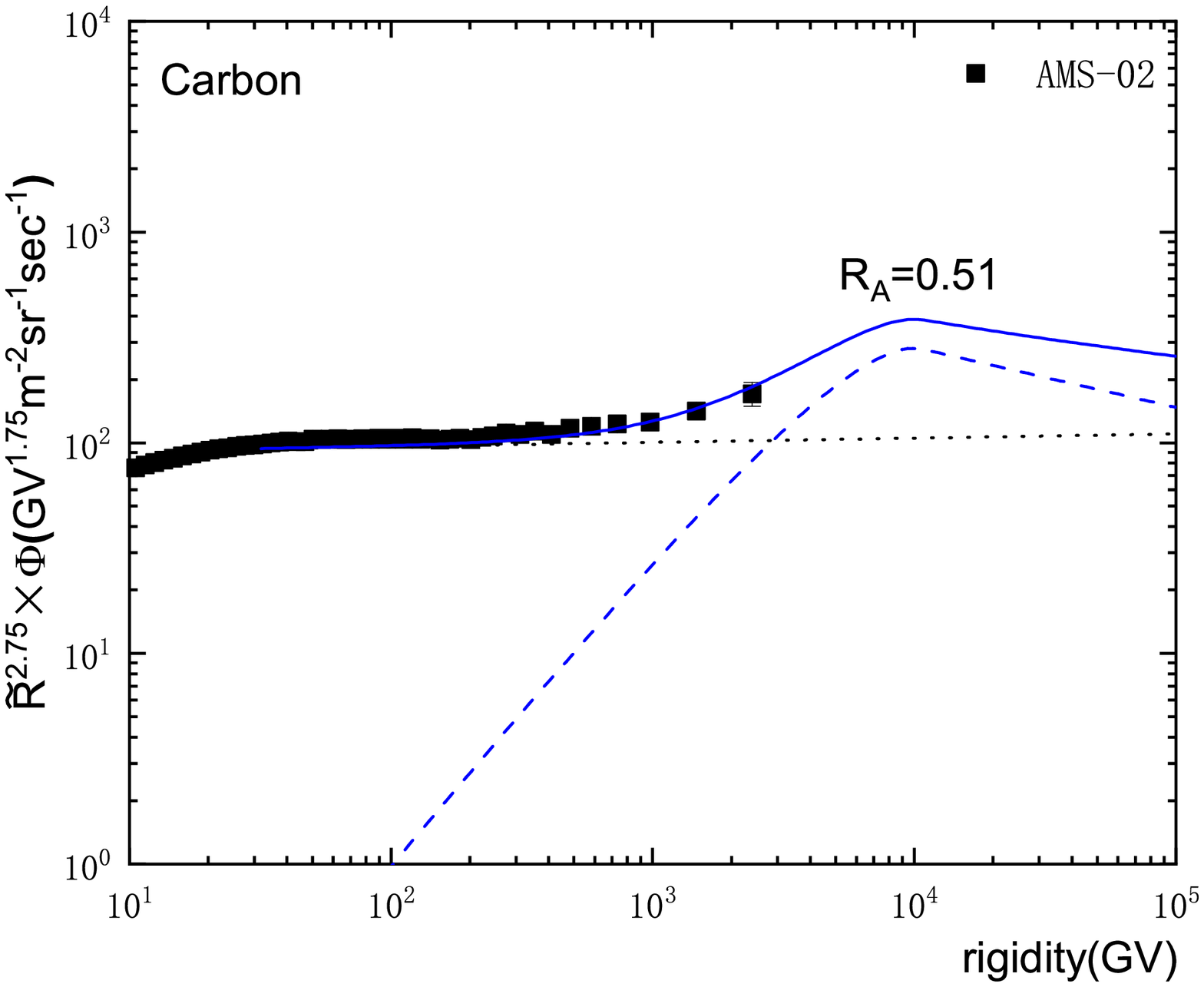}
        \caption{\label{fig:fig6} The carbon energy spectrum multiplied by
$\tilde{R}^{2.75}$ as function of rigidity (solid curve). The dotted
curve is background. The GC-contributions (dashed curve) are
predicted by Equation (4.4). The $\tilde{R}$ position of the peak is
fixed by the positron spectrum. A free parameter $R_A=0.51$. Data
are taken from the AMS-02 [29].}\label{Fig.2}

    \end{center}
\end{figure}

\begin{figure}
    \begin{center}

 \includegraphics[width=0.8\textwidth]{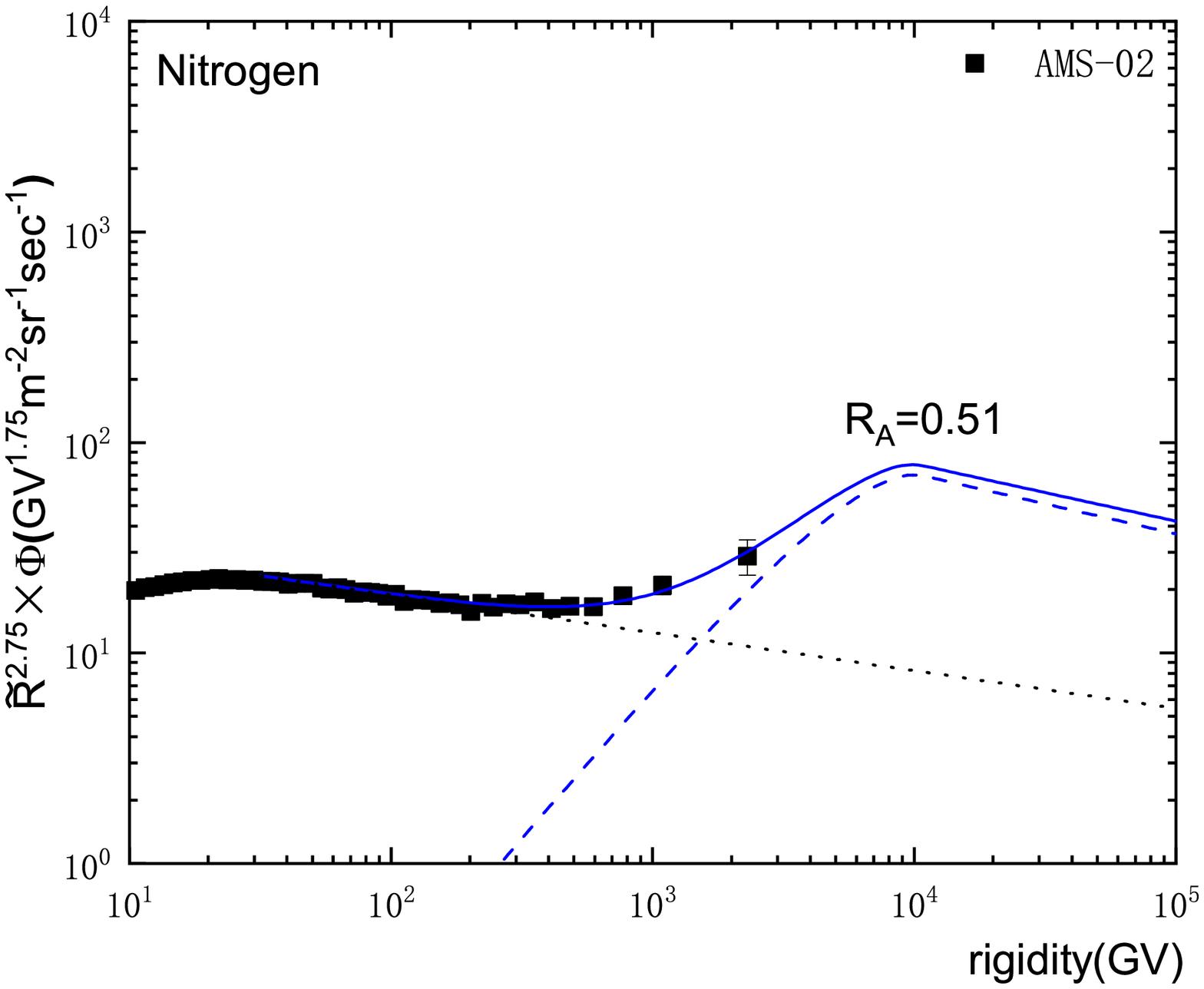}
        \caption{\label{fig:fig7} The nitrogen energy spectrum multiplied by
$\tilde{R}^{2.75}$ as function of rigidity (solid curve). The dotted
curve is background. The GC-contributions (dashed curve) are
predicted by Equation (4.4). The  $\tilde{R}$ position of the peak
is fixed by the positron spectrum. A free parameter $R_A=0.51$. Data
are taken from the AMS-02 [29].}\label{Fig.7}

    \end{center}
\end{figure}

\begin{figure}
    \begin{center}

 \includegraphics[width=0.8\textwidth]{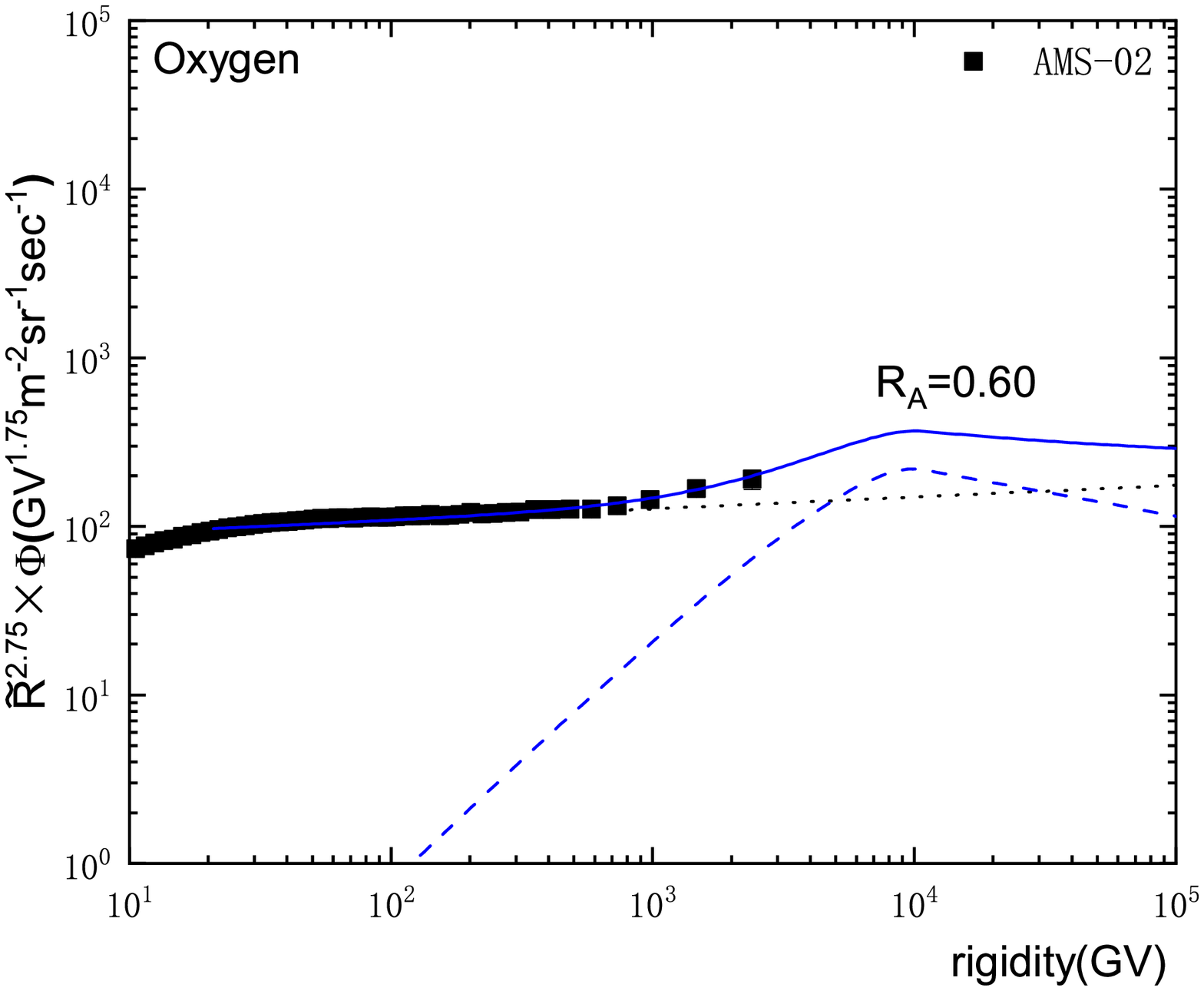}
        \caption{\label{fig:fig8} The oxygen energy spectrum multiplied by
$\tilde{R}^{2.75}$ as function of rigidity (solid curve). The dotted
curve is background. The GC-contributions (dashed curve) are
predicted by Equation (4.4). The position $\tilde{R}$ of the peak is
fixed by the positron spectrum. A free parameter $R_A=0.60$. Data
are taken from the AMS-02 [29].}\label{Fig.8}

    \end{center}
\end{figure}

\begin{figure}
    \begin{center}

 \includegraphics[width=0.8\textwidth]{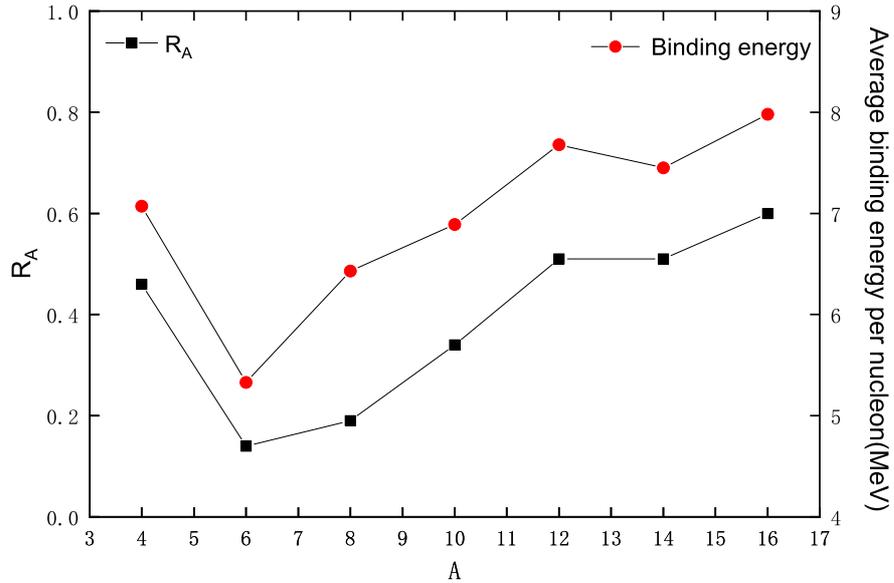}
        \caption{\label{fig:fig9} Values of parameter $R_A$ in Equation (4.5)
    for He, Li, Be, B, C, N and O, and corresponding average binding energy per nucleon. }\label{Fig.9}

    \end{center}
\end{figure}

\end{document}